\documentclass[onecolumn]{aastex7}

\defcitealias{Lu_TRACE}{L24}
\newcommand{\cta}{\citetalias}

\usepackage{url,soul}

\begin{document}

\title{Collisional Fragmentation Support in TRACE}

\author[0000-0003-0834-8645]{Tiger Lu}
\affiliation{Department of Astronomy, Yale University, New Haven, CT 06511, USA}
\email[show]{tiger.lu@yale.edu}  

\author[0009-0006-9796-3568]{Haniyeh Tajer}
\affiliation{School of Earth Sciences, The Ohio State University, 125 South Oval Mall, Columbus OH, 43210, USA}
\email{tajer.1@osu.edu}

\author[0000-0003-0834-8645]{David M. Hernandez}
\affiliation{Department of Astronomy, Yale University, New Haven, CT 06511, USA}
\email{david.m.hernandez@yale.edu}  

\author[0000-0003-1927-731X]{Hanno Rein} 
\affiliation{Department of Physical and Environmental Sciences, University of Toronto at Scarborough, Toronto, Ontario M1C 1A4, Canada}
\affiliation{David A. Dunlap Department of Astronomy and Astrophysics, University of Toronto, Toronto, Ontario, M5S 3H4, Canada}
\email{hanno.rein@utoronto.ca}

\author[0009-0007-9211-2884]{Yurou Liu} 
\affiliation{Department of Astronomy, Yale University, New Haven, CT 06511, USA}
\email{yurou.liu@yale.edu}

\author[0000-0002-7670-670X]{Malena Rice}
\affiliation{Department of Astronomy, Yale University, New Haven, CT 06511, USA}
\email{malena.rice@yale.edu}

%%%%%%%%%%%%%%%%%%%%%%%%%%%%%%%%%%%%%%
\begin{abstract}

We present improved collision support for \texttt{TRACE}, a state-of-the-art hybrid integrator in \texttt{REBOUND}. \texttt{TRACE} now supports collisional fragmentation and can handle both removing and adding particles mid-timestep. We describe the back-end logic implemented for robust collision support, and compare \texttt{TRACE}'s performance to other integrators including \texttt{MERCURIUS} on a large-\textit{N} protoplanetary disk simulation with various collision prescriptions, a system which \texttt{TRACE} previously could not handle. \texttt{TRACE} matches the behavior of these integrators, while offering potentially vast speedups of over $70$x. All updates described in this Note are available with the most recent public release of \texttt{REBOUND}.

\end{abstract}
%%%%%%%%%%%%%%%%%%%%%%%%%%%%%%%%%%%%%%%%%%%
%%%%%%%%%%%%%%%%%%%%%%%%%%%%%%%%%%%%%%%%%%%%%%%%
\section{Introduction}
\texttt{TRACE} \citep[][hereafter L24]{Lu_TRACE} is a novel time-reversible hybrid integrator for the planetary \textit{N}-body problem, based on the methods of \cite{hernandez_2023} and implemented in the open-source \texttt{REBOUND} package \citep{Rein_2012}. Designed as the successor to the \texttt{MERCURIUS} integrator \citep{rein_2019}, \texttt{TRACE} surpasses \texttt{MERCURIUS} in both speed and accuracy and is the recommended integrator within the \texttt{REBOUND} environment for chaotic systems where close encounters occur and where exact trajectories are not required.

While \texttt{TRACE} was rigorously tested for pure \textit{N}-body dynamics in \cta{Lu_TRACE}, its initial public release offered only basic collision handling, limited to detecting collisions via instant overlapping physical radii and modeling perfect mergers. Notably, \texttt{TRACE} lacked support for collisional fragmentation \citep[e.g.][]{leinhardt2012collisions, childs2022collisional} or any collision model that involved generating additional particles — features which are essential for studying processes such as planetesimal and planet formation. While collisions inherently break both time-reversibility and symplecticity \citep[e.g.][]{hairer2006geometric}, in practice these errors typically do not qualitatively affect system evolution.

In this Note, we detail improved collision support for large-\textit{N} systems to the \texttt{TRACE} algorithm. \texttt{TRACE} now supports all collision resolution routines (including arbitrary user-defined routines). We also test its performance on a large-\textit{N} system with a realistic fragmentation prescription.

\section{Collision Logic}
We first briefly review the \texttt{TRACE} algorithm. For an in-depth explanation, see Section 4 of \cta{Lu_TRACE}. For one \texttt{TRACE} timestep:

\begin{enumerate}
    \item Every pair of particles is checked for close encounters, a mutual separation criterion informed by each particle's Hill radius by default.

    \item Particles flagged as undergoing close encounters are integrated forward in time with either \texttt{BS} \citep{press02, Lu_2023} or \texttt{IAS15} \citep{rein_ias15, pham2024new}. Particles that are not flagged for close encounter are integrated forward in time assuming perturbed Keplerian motion using \texttt{WHFast} \citep{rein_2015}.

    \item Every pair of particles is checked again for close encounters. If any pair of particles initially not in a close encounter is in one after the timestep, the step is rejected. We then re-run the step, demanding that this particle pair is integrated with the close encounter prescription (\texttt{BS} or \texttt{IAS15}). This step ensures time-reversibility.
\end{enumerate}
Collisions with particles added/removed mid-timestep are naturally incompatible with Step (3) -- they generate/remove particle-particle pairs that did not exist pre-timestep/now cannot be compared with the associated pre-timestep pair. In \cta{Lu_TRACE}, only collisions that maintained or decreased the number of particles in the simulation were supported. The new handling of collisions is as follows:

\begin{enumerate}
    \item During a close encounter, \texttt{TRACE} checks for physical collisions every \texttt{BS} or \texttt{IAS15} substep. If a collision is detected, the timestep is flagged as irreversible and automatically accepted.

    \item If resolving the detected collision generates new particles, every new particle is assumed to be in a close encounter with both a) every particle flagged for close encounter in Step (1) and b) every other new particle generated in the collision. All new particles, along with the particles previously flagged for close encounters, are integrated forward in time with either \texttt{BS} or \texttt{IAS15}.
\end{enumerate}

\section{Performance Testing}
\label{sec:testing}
We now describe a test of the new collision support on a large-\textit{N} system described in \citep{Chambers_1999}. 30 planetary embryos are initialized around a sun-like star with nearly circular, coplanar orbits. The only difference between our tests and those of \cite{Chambers_1999} is that our disk is closer to the host star (with semimajor axes ranging from $0.1$ to $0.5$ AU) to induce more collisions. We integrate for $10^5$ years, comparing the performance of \texttt{TRACE}, \texttt{MERCURIUS}, \texttt{IAS15}, and \texttt{BS}. For \texttt{TRACE} and \texttt{MERCURIUS}, we adopt a global timestep -- the timestep taken by the Wisdom-Holman component -- of $5$ days. We turn off the pericenter switching condition in \texttt{TRACE} for these tests, and set the critical switching radius to $3.63$ Hill radii. Both are chosen for a better direct comparison with \texttt{MERCURIUS}.

We tested three simple collision prescriptions. Two of these -- perfect mergers and hard-sphere collisions -- were already implemented in \texttt{REBOUND}. We also wrote a simple fragmentation prescription, where two colliding particles generate one large fragment moving in the direction of the original center of mass, and two small fragments moving in the opposite direction. Collision detection is always handled via instantly overlapping physical radii, \texttt{REB\_COLLISION\_DIRECT}. While this is not a fully physical collision outcome, this relatively simple prescription allows for better comparison across integrators.

\begin{figure*}
    \centering
    \includegraphics{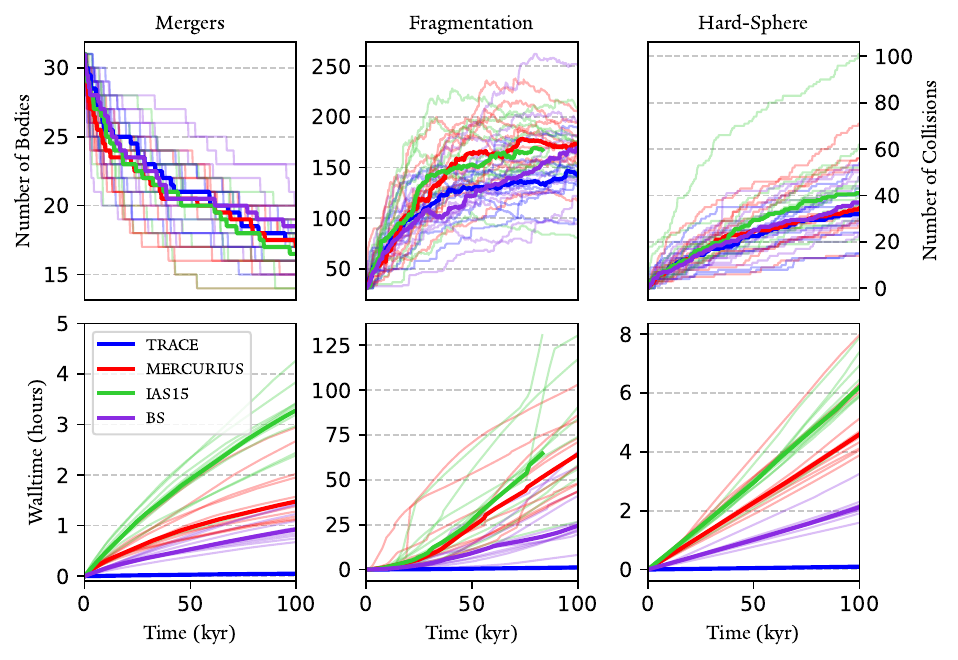}
    \caption{Comparison of \texttt{TRACE} (blue), \texttt{MERCURIUS} (red), \texttt{IAS15} (green) and \texttt{BS} (violet) for a protoplanetary disk simulation inspired by \cite{Chambers_1999} with three different collision prescriptions -- perfect mergers (left), simple fragmentation (center), and perfectly elastic hard-sphere collisions (right). On the top row, we track surviving number of particles (mergers \& fragmentation) and total number of collisions (hard-sphere). On the bottom row, we track runtime. We run each integrator on ten random realizations of the initial conditions. Each individual run is plotted as a thin translucent line. The median of all runs for each integrator is plotted in a thicker solid line. One of the \texttt{IAS15} runs proved too computationally expensive to run to completion.}
    \label{fig:chambersdisk}
\end{figure*}
The results of our simulations are shown in Figure \ref{fig:chambersdisk}. All four integrators behave similarly in a qualitative sense -- the differences between integrators are not statistically significant. In terms of runtime, \texttt{TRACE} is by far the fastest integrator. The median speedup \texttt{TRACE} offers is $36$x/$47$x/$52$x (\texttt{MERCURIUS}), $21$x/$22$x/$22$x (\texttt{BS}) and $56$x/$71$x/$65$x (\texttt{IAS15}) for the merger/fragmentation/hard-sphere prescriptions, respectively.

\section{Conclusion}
In summary, \texttt{TRACE} now supports both removing and adding particles mid-timestep, and thus is able to handle any collision prescription. For such large-\textit{N} systems where realistic collision physics are important, such as planet formation simulations, \texttt{TRACE} yields similar qualitative results to other integrators in the \texttt{REBOUND} ecosystem such as \texttt{MERCURIUS}, while offering potentially over $70$x speedups. We anticipate \texttt{TRACE} to have many useful applications for these communities. 

The updates to \texttt{TRACE} described in this Note are available with the most recent public release of \texttt{REBOUND}. The code used to generate Figure \ref{fig:chambersdisk} of this Note is available at \url{https://github.com/tigerchenlu98/rebound/tree/TRACE_frag_paper/examples/embryos}.

\begin{acknowledgements}
T.L., D.M.H., and M.R. acknowledge support from Heising-Simons Foundation Grant \#2021-2802.
\end{acknowledgements}

\software{\texttt{matplotlib} \citep{hunter_2007},
\texttt{REBOUND} \citep{Rein_2012}}

\bibliography{sample7}{}
\bibliographystyle{aasjournal}

\end{document}